\documentclass[%
reprint,
superscriptaddress,
showpacs,preprintnumbers,
 amsmath,amssymb,
 aps,
prb,
]{revtex4-1}

\usepackage{graphicx}
\usepackage{dcolumn}
\usepackage{bm}

\input{common.tex}

\begin{document}

\preprint{APS/123-QED}

\title{Thickness-dependent magnetic properties and
strain-induced orbital magnetic moment in \SRO\ thin films}%

\author{K. Ishigami}\affiliation{\UTPhys}%
\author{K. Yoshimatsu}\affiliation{\UTChem}%
\author{D. Toyota}\affiliation{\UTChem}
\author{M. Takizawa}\affiliation{\UTPhys}%
\author{T. Yoshida}\affiliation{\UTPhys}%
\author{G. Shibata}\affiliation{\UTPhys}%
\author{T. Harano}\affiliation{\UTPhys}%
\author{Y. Takahashi}\affiliation{\UTPhys}%
\author{T. Kadono}\affiliation{\UTPhys}%
\author{V. K. Verma}\affiliation{\UTPhys}%
\author{V. R. Singh}\affiliation{\UTPhys}%

\author{Y. Takeda}\affiliation{\JAEA}%
\author{T. Okane}\affiliation{\JAEA}%
\author{Y. Saitoh}\affiliation{\JAEA}%
\author{H. Yamagami}\affiliation{\JAEA}%

\author{T. Koide}\affiliation{\PFIMSS}%
\author{M. Oshima}\affiliation{\UTChem}%
\author{H. Kumigashira}\affiliation{\UTChem}\affiliation{\PFIMSS}%
\author{A. Fujimori}\affiliation{\UTPhys}\affiliation{\UTCmpS}\affiliation{\JAEA}%

\date{\today}

\begin{abstract}

Thin films of the ferromagnetic metal \SRO\ (SRO) show a
varying easy magnetization axis depending on the epitaxial
strain and undergo a metal-to-insulator transition with
decreasing film thickness.  We have investigated the
magnetic properties of SRO thin films with varying
thicknesses fabricated on \STO(001) substrates by soft x-ray
magnetic circular dichroism (XMCD) at the Ru M$_{2,3}$ edge.
Results have shown that, with decreasing film thickness, the
film changes from ferromagnetic to non-magnetic around
\SI{3}{monolayer} thickness, consistent with previous
magnetization and magneto-optical Kerr effect measurements.
The orbital magnetic moment perpendicular to the film was
found to be $\sim$~\SI{0.1}{\si{\muB\per Ru atom}}, and
remained nearly unchanged with decreasing film thickness
while the spin magnetic moment decreases. Mechanism for the
formation of the orbital magnetic moment is discussed based
on the electronic structure of the compressively strained
SRO film.

\end{abstract}

\maketitle

\section{Introduction}

\SRO\ (SRO), a 4\orbit{d} transition metal oxide with the
perovskite-type structure, is a ferromagnetic metal with a
relatively high Curie temperature of $T_c
\sim$~\SI{160}{\kelvin}. The electrical resistivity does not
saturate even above \SI{500}{\kelvin}, where the Ioffe-Regel
limit is exceeded
\cite{ref0256-307X-23-8-072,ref0953-8984-14-31-310},
indicating highly incoherent nature of the metallic state,
i.e., so-called a ``bad metallic'' behavior. From the device
application point of view, SRO is a promising material,
e.g., as electrodes because of the chemical stability and
the structural compatibility with many functional oxides.

It has been known that the electronic and magnetic
properties of epitaxially grown thin films are profoundly
affected by the film thickness  and the epitaxial strain
from the substrates. Several studies have shown that, with
decreasing film thickness, SRO thin films exhibit a
metal-to-insulator transition and concomitant loss of
ferromagnetism at a critical thickness of several monolayer
(ML)
\cite{toyota:162508,Toyota2006,ref10.1103/PhysRevB.79.140407}.
Using the laser molecular beam epitaxy (MBE) method, Toyota
\etal \cite{toyota:162508,Toyota2006} reported the
thickness-dependent electronic structure of SRO films grown
on Nb-doped \STO(001) (Nb:STO) substrates by measuring the
electrical resistivity and valence-band photoemission
spectra. They showed from the temperature dependence of the
resistivity that the films changed from metallic to
insulating with decreasing thickness. The photoemission
spectra of the SRO thin films showed a clear Fermi edge for
film thicknesses above \SI{5}{ML}.  With decreasing film
thickness, the center of the Ru 4\orbit{d} band moved
towards higher binding energies and the intensity at the \EF
decreased, resulting in an energy gap opening at the \EF
below \SI{4}{ML}. This indicates that SRO thin film
undergoes a metal-insulator transition between \SI{4}{ML}
and \SI{5}{ML} thicknesses, consistent with the resistivity
measurements.  The magnetic properties of SRO thin films
grown on STO(001) substrates have been investigated by Xia
\etal through magneto-optical Kerr effect measurements
\cite{ref10.1103/PhysRevB.79.140407}.  With decreasing film
thickness, the film showed a transition from ferromagnetic
to paramagnetic between \SI{4}{ML} and \SI{3}{ML}.
Mahadevan \etal \cite{ref10.1103/PhysRevB.80.035106}
performed a density-functional calculation, and found that
the SRO film indeed exhibits a thickness-dependent
transition from a ferromagnetic metal to an
antiferromagnetic insulator at \SI{4}{ML}.  As for the
magnetic anisotropy, the magnetic moment was found nearly
perpendicular to the film surface.  With increasing in-plane
lattice constant through increasing the Ba content in \BSTO
(BSTO) substrates, the easy magnetization axis changed from
out-of-plane to in-plane \cite{JJAP.43.L227}.

It has been generally considered that perpendicular magnetic
anisotropy arises from magneto-crystalline anisotropy caused
by spin-orbit interaction. Bruno has shown that the MCA
energy is proportional to the difference in the orbital
magnetic moment between the perpendicular and in-plane
directions\cite{PhysRevB.39.865}, and this has been
confirmed for 3\orbit{d} transition metals such as
Au/Co/Au(111) thin films\cite{PhysRevLett.75.3752} and
FeCo/Ni multilayers\cite{PhysRevB.87.054416}. If the Bruno
theory is applicable to Ru compounds, too, the SRO thin
films grown on STO are expected to exhibit a finite orbital
magnetic moment perpendicular to the plane although the
orbital magnetic moment in bulk SRO has been reported to be
negligibly small\cite{ref10.1103/PhysRevB.76.184441}. So
far, different values have been reported for the orbital
magnetic moment of Ru in SRO thin films grown on STO
substrates not only with (001) surfaces but also with (110)
and (111)
surfaces\cite{PhysRevB.85.134429,PhysRevB.91.075127}
and the issue still remains controversial.

The purpose of the present study is to elucidate the
thickness-dependent magnetic properties of the SRO thin
films grown on STO(001) substrates through the measurements
of the spin and orbital magnetic moments using \XMCD (XMCD).
We indeed observed a finite orbital magnetic moment of
$\sim$~\SI{0.1}{\si{\muB\per Ru}} atom perpendicular to the
film surface.  The origin of the perpendicular orbital
magnetic moment, which should be related to the
perpendicular magnetic anisotropy according to
Bruno\cite{PhysRevB.39.865}, shall be discussed.

\section{Experiment}

SRO thin films were fabricated on \TiOtwo-terminated
\SI{0.05}{\%} Nb-doped STO(001) substrates by the laser-MBE
method with precise control of thickness. The wet-etched
STO(001) substrates with \TiOtwo-termination were annealed
at \SI{1100}{\celsius} for 2 hours under an oxygen pressure
of \SI{1e-7}{Torr} to ensure atomically flat surfaces.
Sintered SRO pellets were used as targets.  A Nd:YAG
(yttrium aluminum garnet) laser was used for ablation in its
frequency-tripled mode ($\lambda$ = \SI{355}{\nano\meter})
at a repetition rate of 1 Hz. During the deposition, the
substrate temperature was kept at \SI{750}{\celsius} and the
oxygen pressure at \SI{1e-3}{Torr}. The thickness of thin
films were determined by reflection high-energy
electron-diffraction (RHEED) oscillation.  The RHEED pattern
showed Kikuchi lines and no three-dimensional Bragg spots,
which means that the SRO thin films have flatter surfaces
and are better crystallized than those fabricated in
previous works\cite{toyota:162508,Toyota2006}.  \Exsitu
atomic force microscope studies showed step-and-terrace
structures for all the samples. For the samples with
4-\SI{8}{ML} thickness, however, the step edges were
irregular, which means that the step-flow growth condition
was not achieved \cite{DtoyotaMasterThesis}. The
pseudo-cubic lattice constant of SRO is
$\sim$~\SI{3.92}{\AA} and is larger than the lattice
constant \SI{3.905}{\AA} of STO by \SI{0.4}{\%}, meaning
that the SRO thin films grown on STO substrates are under
compressive strain.

Soft x-ray photoemission measurements were performed at the
undulator beamline BL-2C of Photon Factory, KEK. \XAS (XAS)
and XMCD measurements were performed at the helical
undulator beamline BL23SU of SPring-8 except for the sample
of \SI{50}{ML} thickness. As for the sample of  \SI{50}{ML}
thickness, XAS and XMCD measurements were performed at the
undulator beamline BL-16A of Photon Factory, KEK . For the
samples measured at SPring-8, in order to eliminate spurious
signals in XMCD spectra, the helicity of the incident
circularly polarized light was switched at each photon
energy, and two XMCD spectra obtained using opposite
magnetic-field directions were averaged.  The Ru
M$_{2,3}$-edge (Ru 3\orbit{p}$\to$4\orbit{d}) XAS and XMCD
spectra were taken at \SI{20}{\kelvin} by the total electron
yield mode with negative bias. External static magnetic
field of 0.1-\SI{5}{T} was applied perpendicular to the film
surfaces.

\section{Result and Discussion}

Magnetization-temperature curves of the SRO films with
various thicknesses were measured using a superconducting
quantum interference device (SQUID) and are shown in
\figr{SRO-MT}. They show that the magnetization quickly
increases above \SI{4}{ML}, indicating that a
paramagnetic-ferromagnetic transition occurs between 3 and
\SI{4}{ML} thicknesses.

Photoemission spectra in the valence-band region are shown
in \figr{SRO-xML-XPS}(a). One can see three structures
originating from the O 2\orbit{p} band, one of which is
located around \SI{4}{\eV} and the others are located around
\SI{7}{\eV} and \SI{8}{\eV}\cite{PhysRevB.60.2281,
:/content/aip/journal/jap/96/12/10.1063/1.1814175}.
Photoemission within $\sim$~\SI{2}{\eV} of \EF is originated
from the Ru 4\orbit{d}
band\cite{:/content/aip/journal/jap/96/12/10.1063/1.1814175}.
The spectrum for the film thicknesses of \SI{2}{ML} exhibits
an energy gap at \EF, as clearly seen in the spectra near
\EF [\figr{SRO-xML-XPS}(b)]. The leading edge of the Ru
4\orbit{d} band reaches \EF at \SI{3}{ML} and the Fermi edge
is established at \SI{4}{ML}, indicating a
thickness-dependent insulator-to-metal transition at a
critical film thickness between 3 and \SI{4}{ML}.  This
critical thickness is the same as that reported in Ref.
\citenum{ref10.1103/PhysRevB.79.140407} but by \SI{1}{ML}
smaller than that reported in Refs.  \citenum{toyota:162508}
and \citenum{Toyota2006}.

\begin{figure}
	\begin{center}
		\includegraphics[width=\linewidth]{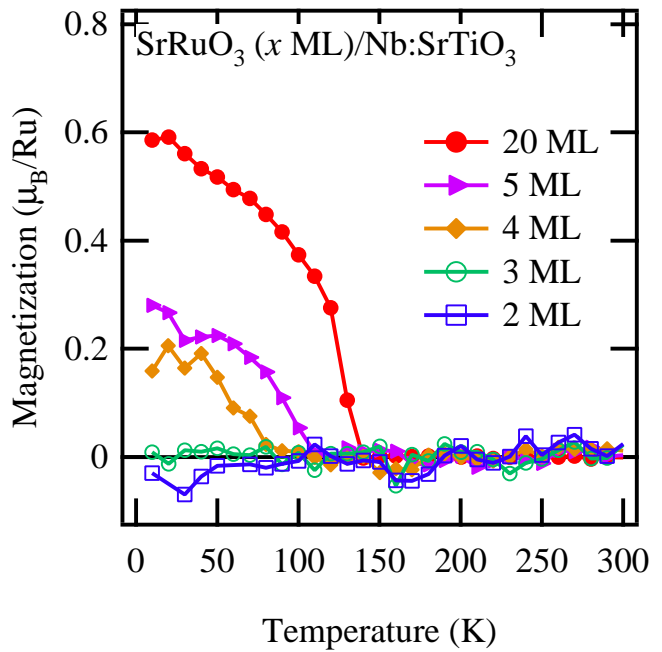}
		\caption{\CO
		Temperature
		dependence of the magnetization of \SRO\ thin
		films grown on Nb:STO substrates with
		various thicknesses measured by remnant
		magnetization after field cooling at
		$\mu_\text{0}H$ = \SI{3}{T}. The films with
		thicknesses greater than \SI{4}{ML} show
		ferromagnetic behavior.}
		\figl{SRO-MT}
	\end{center}
\end{figure}

\begin{figure}
	\begin{center}
		\includegraphics[width=\linewidth]{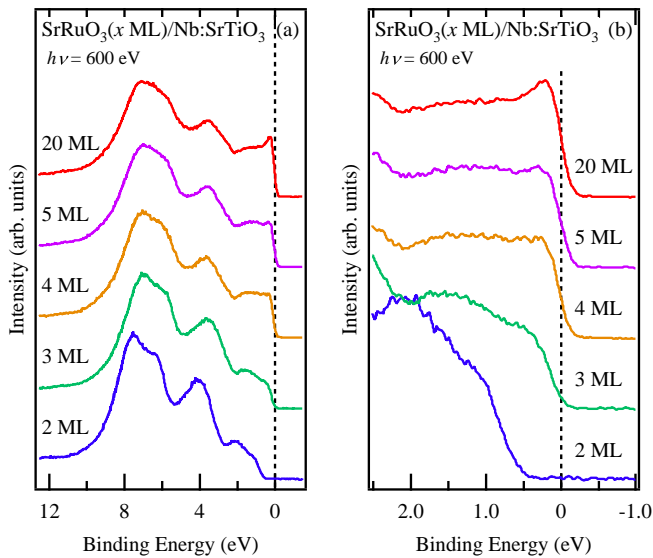}
		\caption{\CO Thickness
		dependence of the \insitu valence-band
		photoemission spectra of \SRO\ thin films
		grown on Nb-doped \STO\ substrates. (a) The
		entire valence band, and (b) near \EF
		region. The Fermi cutoff is clearly seen
		above \SI{3}{ML}\cite{DtoyotaMasterThesis},
		indicating the metallic nature of the films.}
		\figl{SRO-xML-XPS}
	\end{center}
\end{figure}

\figurer{SRO-MCD-XAS-xML} shows the Ru M$_{2,3}$ XAS and
XMCD spectra of the \SI{50}{ML} and \SI{4}{ML} SRO films at
the magnetic field of $\mu_\text{0}H$ = \SI{5.0}{T}. For the
\SI{50}{ML}-thick SRO film [\figr{SRO-MCD-XAS-xML}(a)],
clear Ru M$_{2,3}$ XAS and XMCD spectra were observed. For
the \SI{4}{ML}-thick film [\figr{SRO-MCD-XAS-xML}(b)], the
strong Ti L$_{2,3}$-derived peaks from the STO substrate
overlap the Ru M$_3$ (3\orbit{p}$_{3/2}$$\to$4\orbit{d})
peak because the SRO thickness of \SI{4}{ML} was not thick
enough compared with the probing depth of XAS. On the other
hand, the Ru M$_2$ (3\orbit{p}$_{1/2}$$\to$4\orbit{d}) edge
at \SI{484.4}{\eV} does not overlap the Ti L$_{2,3}$ edge,
and is therefore better resolved. The Ru M$_3$ edge XAS
buried under the Ti L$_{2,3}$ XAS is deduced from the Ru
M$_2$ peak intensity, and is plotted, by a dashed curve in
\figr{SRO-MCD-XAS-xML}(b). Taking the difference between the
XAS spectra for right and left circularly polarized light,
we have obtained the XMCD spectra as shown in the bottom
panels of Figs. \ref{fig:SRO-MCD-XAS-xML}(a) and (b). In
\figr{SRO-MCD-XAS-xML}(b), despite the strong XAS signals
from the Ti L$_{2,3}$ edge, no spurious XMCD signals due to
the Ti L$_{2,3}$ XAS are detected. Since the XMCD spectrum
in \figr{SRO-MCD-XAS-xML}(b) was measured by reversing the
photon helicity at each photon energy and also by reversing
the magnetic field, we consider that the intrinsic Ru
M$_{2,3}$ XMCD of SRO was clearly observed.

\begin{figure}
	\begin{center}
		\includegraphics[width=\linewidth]{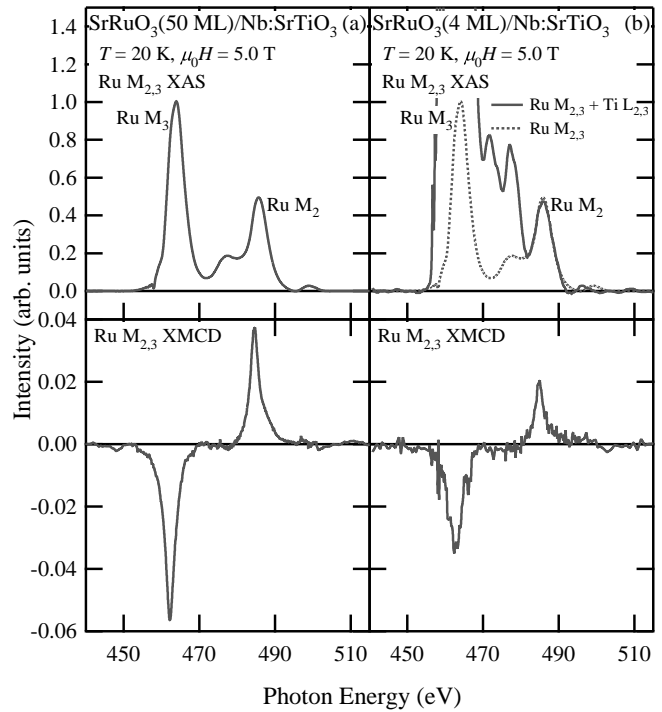}
		\caption{\CO Ru M$_{2,3}$-edge XAS and XMCD
		spectra of \SRO\ thin films with the thicknesses
		of \SI{50}{ML} (a) and \SI{4}{ML} (b).
		The dashed curves in panel (b)
		is the XAS spectrum of the \SI{50}{ML} film
		plotted so that the Ru M$_{2}$ intensities
		coincide.}
		\figl{SRO-MCD-XAS-xML}
	\end{center}
\end{figure}

\begin{figure}
	\begin{center}
		\includegraphics[width=\linewidth]{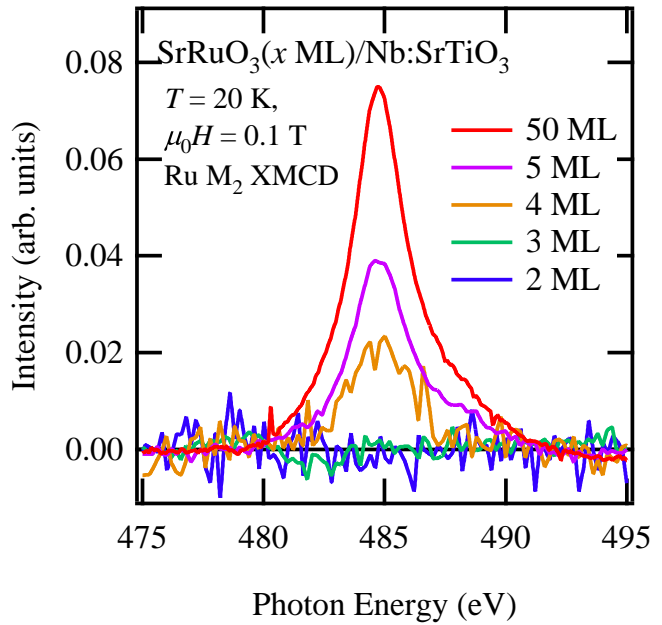}
		\caption{\CO Thickness dependence of the Ru
		M$_2$-edge XMCD spectra of \SRO\ thin films.
		The intensities have been normalized to the
		Ru M$_2$-edge XAS intensity.}
		\figl{SRO-MCD-XAS-Int-xML-xT}
	\end{center}
\end{figure}

\figurer{SRO-MCD-XAS-Int-xML-xT} shows thus obtained Ru
M$_{2}$-edge XMCD spectra of SRO films with various
thicknesses taken at the low magnetic field of
$\mu_\text{0}H$ = \SI{0.1}{T}. In such a low magnetic field,
while ferromagnetic samples show strong XMCD signals
paramagnetic samples show only very weak XMCD signals. One
can see that the XMCD intensity decreases with decreasing
film thickness and vanishes at \SI{3}{ML}, signaling a
ferromagnetic-to-paramagnetic transition between \SI{4}{ML}
and \SI{3}{ML}. The thickness and magnetic-field dependences
of the Ru M$_2$-edge XMCD intensity were measured and are
summarized in \figr{SRO-MCDratio-xML-xT}.  According to
\figr{SRO-MCDratio-xML-xT}(a), the XMCD intensity of the
\SI{3}{ML} film show an increase at high magnetic fields,
indicating a paramagnetic (or an antiferromagnetic) ground
state and a possible metamagnetic behavior. The XMCD
intensities of the \SI{4}{ML} and thicker films, on the
other hand, show an abrupt increase from $\mu_\text{0}H$ =
\SI{0}{T} to almost saturated values at $\mu_\text{0}H$ =
\SI{0.1}{T}, confirming that these films are ferromagnetic.
\figurer{SRO-MCDratio-xML-xT}(b) is the thickness dependence
of XMCD intensities at several fixed magnetic fields. They
all show increase with film thickness above \SI{4}{ML} at
all applied fields.

The thickness and magnetic-field dependences of the orbital
and spin magnetic moments have been derived using the XMCD
sum rules \cite{ref10.1103/PhysRevLett.68.1943,
PhysRevLett.70.694} and are plotted in
\figr{SRO-Moment-xML-xT}.  The spin magnetic moment of the
thick SRO films is comparable to that of bulk SRO
sample\cite{ref10.1103/PhysRevB.76.184441} ($m_\text{spin}
\simeq$~\SI{0.6}{\si{\muB\per Ru}}). This does not follow
the result of first-principles calculation on SRO under
epitaxial strain which indicates that the magnetic moment
should decrease by $\sim$~\SI{10}{\%} under the compressive
strain of \SI{0.4}{\%} from the STO
substrate\cite{PhysRevB.77.214410}, probably because the
accuracy of the previous XMCD measurements on bulk SRO
crystal \cite{ref10.1103/PhysRevB.76.184441} was not
sufficient to discuss the subtle differences between the
bulk and thin film data.

The orbital magnetic moment ($m_\text{orb}
\simeq$~\SI{0.08}{\si{\muB\per Ru}}) of the thick SRO film
is much smaller than 0.2-\SI{0.3}{\muB} reported in the
previous Ru M$_{2,3}$ XMCD study of SRO films grown on
STO(001) and (111) substrates\cite{PhysRevB.85.134429} but
significantly larger than that ($m_\text{orb}
\simeq$~0.0-\SI{0.03}{\si{\muB\per Ru}}) reported by the
very recent XMCD study at the Ru L$_{2,3}$ edge XMCD of SRO
films grown on STO(001) and (111)\cite{PhysRevB.91.075127}.
The discrepancy between the Ru L$_{2,3}$ edge and the
present M$_{2,3}$ edge studies even without overlapping Ti
L$_{2,3}$ edges may be due to the large
($\sim$~\SI{130}{\eV}) spin-orbit splitting of the Ru
L$_{2,3}$ edge which may make the transition-matrix elements
for the L$_2$ and L$_3$ edges slightly different.  Because
the XMCD sum rules have been derived under the assumption
that the radial part of the core-level wave functions is the
same for the $j=l+1/2$ and $l-1/2$ core levels, it may be
different if the spin-orbit interaction is very strong. If
the L$_2$ edge has a larger (smaller) matrix element than
L$_3$, the orbital magnetic moment will be underestimated
(overestimated).  The reason why the $m_\text{orb}$ value
reported in Ref. \citenum{PhysRevB.85.134429} is much larger
than ours is not known at present. This discrepancy may be
related to the unusually large $m_\text{spin}$ value
($\sim$~\SI{3.4}{\si{\muB\per Ru}}) reported in Ref.
\citenum{PhysRevB.85.134429} compared to ours
($\sim$~\SI{0.6}{\si{\muB\per Ru}}) as well as to the value
deduced from bulk magnetization measurements
($\sim$~\SI{1.0}{\si{\muB\per Ru}})
\cite{ref10.1103/PhysRevB.76.184441}.  As for the \SI{4}{ML}
film, since the spin magnetic moment is smaller
($m_\text{spin} \simeq$~\SI{0.4}{\si{\muB\per Ru}}) than
that of bulk SRO, the ratio
$m_\text{orb}/(m_\text{spin}+7m_T)$ is larger in the thin
film than in the \SI{50}{ML} thick film by a factor of
$\sim$~2, as plotted in \figr{SRO-Moment-xML-xT}.

The finite orbital magnetic moment perpendicular to the film
can be understood from the band structure of SRO as follows:
In SRO, the $t_{2g}$ band is partially occupied and
spin-polarized while the $e_g$ band is empty. Under the
compressive strain, the $t_{2g}$ band is split into the
wider $d_{xy}$ band and the narrower doubly degenerate
$d_{yz}$/$d_{zx}$ bands. When the spins are perpendicular to
the film, i.e., along the $z$ direction, the $d_{yz}$ and
$d_{zx}$ bands are mixed through (the $L_{z}$$S_{z}$ term
of) the spin-orbit interaction, and the orbital magnetic
moment along the $z$ direction is induced. When the spins
are parallel to the film, e.g., along the $x$ direction, the
$d_{zx}$ and $d_{xy}$ bands are mixed through (the
$L_{x}$$S_{x}$ term of) the spin-orbit interaction, and the
orbital magnetic moment is induced along the $x$ direction,
however, the induced orbital moment is smaller because the
wider $d_{xy}$ band is
involved.\cite{SRO_Thickness_Dependence_XMCD_comment00}
According to Bruno\cite{PhysRevB.39.865}, the larger orbital
magnetic moment perpendicular to the film  than that
parallel to it should lead to the perpendicular magnetic
anisotropy, as confirmed by XMCD for several systems
including Co thin films sandwiched by
Au(111).\cite{PhysRevLett.75.3752} For the Au/Co/Au(111)
film, the orbital magnetic moment perpendicular to the film
increases with decreasing Co film
thickness.\cite{PhysRevLett.75.3752} In the case of the SRO
thin films, the increase of the ratio
$m_\text{orb}/(m_\text{spin}+7m_T)$ with decreasing film
thickness may be induced by a similar mechanism to the
Au/Co/Au film. In order to see whether Bruno's
theory\cite{PhysRevB.39.865} holds or not for the SRO films,
the orbital magnetic moment parallel to the film as well as
the orbital magnetic moment of SRO thin films grown on
substrates having different lattice constants such as
BSTO\cite{JJAP.43.L227} remain to be measured in the future.
On the theoretical side, first-principles calculation on SRO
thin films explicitly including the Ru 4\orbit{d} spin-orbit
coupling ($\sim$~\SI{150}{\milli\eV}), which is larger than
that of Co 3\orbit{d} ($\sim$~\SI{70}{\milli\eV}), is
necessary to quantitatively understand the origin of the
perpendicular orbital magnetic moment and the perpendicular
magnetic anisotropy.

\begin{figure}
	\begin{center}
		\includegraphics[width=\linewidth]{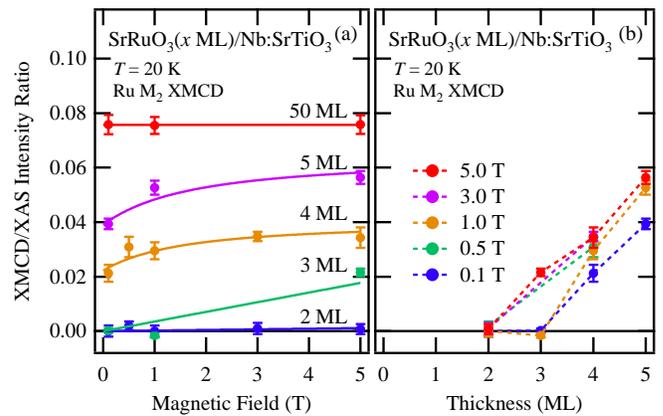}
		\caption{\CO Thickness (a) and
		magnetic-field (b) dependences of the XMCD
		intensities (measured in terms of the
		XMCD/XAS intensity ratio) at the Ru M$_{2}$
		edge of \SRO\ thin films.}
		\figl{SRO-MCDratio-xML-xT}
	\end{center}
\end{figure}

\begin{figure}
	\begin{center}
		\includegraphics[width=\linewidth]{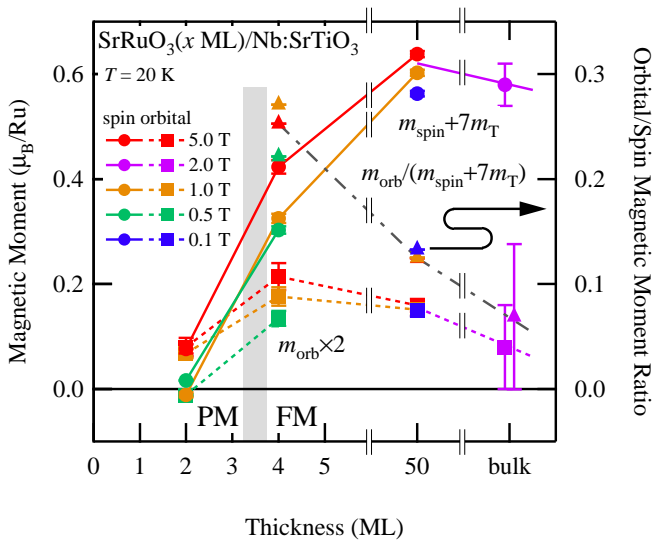}
		\caption{\CO Magnetic-field dependence of
		the (effective) spin magnetic moments
		($m_\text{spin}+7m_\text{T}$) and the
		orbital magnetic moments ($m_\text{orb}$) of
		\SRO\ thin films.  The electron occupation
		number $n_\text{4d}$ is assumed to be 4.
		$m_\text{T}$ is the expected
		value of the magnetic dipole operator which
		originate from the anisotropic distribution
		of the spin density. The data of bulk \SRO\ 
		are taken from Ref.
		\citenum{ref10.1103/PhysRevB.76.184441}.}
 		\figl{SRO-Moment-xML-xT}
	\end{center}
\end{figure}

\section{Summary}

We have performed XMCD measurements on the SRO thin films
with various thicknesses grown on STO(001) substrates. With
decreasing film thickness, the intensity of the XMCD spectra
decreased and the XMCD signal at low magnetic field became
very weak below \SI{3}{ML}, indicating a
ferromagnetic-to-paramagnetic transition. While films with
thicknesses larger than \SI{4}{ML} showed strong,
magnetic-field-independent XMCD, indicating ferromagnetic
behavior, the sample with \SI{3}{ML} thicknesses showed weak
XMCD signals which increases with magnetic field, consistent
with (enhanced) paramagnetic behavior.  The orbital magnetic
moment perpendicular to the film was found to be small but
finite ($\sim$~\SI{0.1}{\si{\muB\per Ru}}).  The origin of
the perpendicular orbital magnetic moment is discussed based
on the band structure of SRO under the compressive strain.

\begin{acknowledgments}
We would like to thank Kenta Amemiya and Masako Sakamaki for
valuable technical support at KEK-PF.  Discussion with P.
Mahadevan is gratefully acknowledged.  This work was
supported by a Grant-in-Aid for Scientific Research from
JSPS (S22224005) and the Quantum Beam Technology Development
Program from JST. The experiment was performed under the
approval of the Photon Factory Program Advisory Committee
(Proposal Nos.  2009G579, 2012G667, 2013S004) and under the
Shared Use Program of JAEA Facilities (Proposal Nos.
2011A3840/BL23SU).
\end{acknowledgments}

\bibliography{all}

\begin{thebibliography}{20}
\bibitem [{SRO()}]{SRO_Thickness_Dependence_XMCD_comment00}%
  \BibitemOpen
  \href@noop {} {}\bibinfo {note} {If the SRO film is grown under tensile
  strain (as in the case of Ref.\citenum{JJAP.43.L227}), the $t_{2g}$ band is
  split into the narrower $d_{xy}$ band and the wider $d_{yz}$/$d_{zx}$ bands,
  and the magnitude of the orbital magnetic moment would be larger for the
  in-plane spin direction, for which the narrower $d_{xy}$ band is involved in
  the spin-orbit coupling.}\BibitemShut {Stop}%
\end{thebibliography}

\end{document}